\documentclass{raa}           

\newcommand{\kms }{\,km\,s$^{-1}$}

\usepackage{graphicx}
\usepackage{txfonts}
\usepackage{natbib}
\usepackage{threeparttable, tablefootnote}
\bibpunct{(}{)}{;}{a}{}{,} 
\usepackage[a4paper=true,pagebackref=true]{hyperref}
\hypersetup{colorlinks = true, linkcolor = green, anchorcolor = red, citecolor = blue, filecolor = red, pagecolor = red, urlcolor = red}
\renewcommand*{\tablefootnote}\textsuperscript{{\alph{footnote}}}
\begin{document}

   \title{Search for H~{\sc i} emission from superdisk candidates associated with radio galaxies}

   \author{Abhijeet Anand\inst{1}
          \and
          Nirupam Roy\inst{2}
          \and
          Gopal-Krishna\inst{3,4}\fnmsep \thanks{Visiting Professor, ARIES}
          }
   \institute{Max-Planck-Institut f{\"u}r Astrophysik, Karl-Schwarzschild-Str. 1, D-85741 Garching, Germany;  {\it abhijeet@mpa-garching.mpg.de}\\
         \and
         Department of Physics, Indian Institute of Science, Bangalore - 560012, India\\
         \and
	     Aryabhatta Research Institute of Observational Sciences (ARIES), Manora Peak, Nainital - 263129, India\\
	 \and
	 Max-Planck-Institut f{\"u}r Radioastronomie, Auf dem H{\"u}gel 69, 53121 Bonn, Germany}
 
  \abstract{Giant gaseous layers (termed "superdisk") have been hypothesized in the past to account for the strip-like radio emission gap (or straight-edged central brightness depression) observed between the twin radio lobes, in over a dozen relatively nearby powerful Fanaroff-Riley Class II radio galaxies. They could also provide a plausible alternative explanation for a range of observations. Although a number of explanations have been proposed for the origin of the superdisks, little is known about their material content. Some X-ray observations of superdisk candidates indicate the presence of hot gas, but a cool dusty medium also seems to be common in them. If they are made of entirely or partly neutral gas, it may be directly detectable and we report here a first attempt to detect/image any neutral hydrogen gas present in the superdisks that are inferred to be present in four nearby radio galaxies. We have not found a positive H{\sc~i} signal in any of the four sources, resulting in tight upper limits on the H{\sc~i} number density in the postulated superdisks, estimated directly from the central rms noise values of the final radio continuum subtracted image. The estimated ranges of the upper limit on neutral hydrogen \textit{number density} and \textit{column density} are $10^{-4}-10^{-3}$ atoms per cm$^3$ and $10^{19}-10^{20}$ atoms per cm$^{2}$, respectively. No positive H{\sc~i} signal is detected even after combining all the four available H{\sc~i} images (with inverse variance weighting). This clearly rules out an H{\sc~i} dominated superdisk as a viable model to explain these structures, however, the possibility of superdisk being made of warm/hot gas still remains open.
   }
   
\titlerunning{Search for H~{\sc i} 21 cm emission from `superdisks'}
\authorrunning{Anand, Roy \& Gopal-Krishna}
   \maketitle
%

\section{Introduction}

Conspicuous radio emission gaps with sharp, quasi-linear edges have been noticed
in the middle of the twin radio lobes of nearly 20 powerful double radio 
sources. These strip-like emission gaps (or sharp-edged brightness depressions)
oriented roughly orthogonally to the axis defined by the radio lobe pair,
have a typical width of about 30 kpc and length 
of at least 50-100 kpc, and are suspected to be caused due to a giant layer 
of thermal gas enveloping the massive elliptical galaxy hosting the double 
radio source \citep[]{GKW1996, GKW2000, GKWJ2007,GKW2009}. As argued in these 
papers, such postulated {\it superdisks} can provide viable alternative 
explanations to a range of well known phenomena associated with radio galaxies
and quasars, namely (i) the radio lobe 
depolarization asymmetry, popularly known as Laing-Garrington effect 
\citep[]{RL1988, GRTN1988}; (ii) correlated radio-optical asymmetry of double 
radio sources \citep[]{MBK1991}; (iii) the occurrence of absorption dips in the 
Lyman-$\alpha$ (Ly-$\alpha$) emission profiles of high-$z$ radio galaxies with 
a total radio extent of up to $\sim 50$ kpc \citep[]{VOR1997,BLW2006}; and (iv) the 
apparent asymmetry of the extended Ly-$\alpha$ emission found associated with the lobes of
high-$z$ radio galaxies (\citealt{GKW2000}, hereafter GKW2000). Since the sharp edges of 
the radio brightness gaps/depressions would only be conspicuous when the axis of the 
double radio source is oriented near the plane of the sky and hence the twin radio lobes
do not appear partially overlapping, the (inferred) 
superdisks probably exist in many more radio galaxies than are actually observed.

Although the postulated superdisks (or, `fat-pancakes') might play a significant 
role in causing several well known correlations exhibited by radio galaxies, as 
mentioned above, progress in understanding their nature is severely hampered 
due to the lack of information about their content. Early hints for the 
presence of dusty material in the superdisks emerged from the observed asymmetry
of the extended Ly-$\alpha$ emission in high-$z$ radio galaxies and from the 
above cited correlated radio-optical asymmetry (GKW2000 and references therein). 
Such a cool component (with dusty clumps) would also be consistent with the 
curious result 
\citep[]{VOR1997,BLW2006} that absorption dips in the Ly-$\alpha$ emission 
profiles of radio galaxies are mostly seen when the overall radio extent of
 the lobes is under $\sim 50$ kpc, i.e., comparable to the typical width of 
the putative superdisks 
\citep[]{GKW1996,GKW2000,GKW2009}. An observation particularly relevant to 
the superdisk hypothesis is the detection of an enormous H~{\sc i} ring (of 
diameter $\sim 200$ kpc) surrounding an early-type galaxy NGC 612 hosting a 
double radio source \citep[][and references therein]{EB2008}. Although a superdisk
signature is not evident in the 
radio images of this particular radio galaxy, 
this could be understood if the aforementioned requirement of the radio axis 
lying close to the sky plane remains unsatisfied in this case.

Motivated by the aforementioned findings and correlations, a number of 
physical scenarios have been 
proposed wherein the superdisks are envisioned to be: \textbf{(a)} A layer of 
dusty neutral and/or ionized gas blocking (at least partially) the backflow of the 
synchrotron plasma of the radio lobes, which is the tidally stretched remnant 
of a disk galaxy captured by the elliptical galaxy that hosts the double radio source 
(\citealt{GKB1997}, see also, \citealt[]{BBLP1992}); \textbf{(b)} A thermal wind 
bubble blown by accretion disk of the active galactic nucleus (AGN), which is 
sandwiched and compressed by the backflowing synchrotron plasma of the twin radio 
lobes \citep[]{GKWJ2007}; \textbf{(c)} A segment of the gaseous filaments of the 
cosmic web in which the galaxy hosting the radio source happens to reside \citep[]{GKW2000,GKW2009}. 
In this scheme, the putative superdisks must have already existed prior to the birth 
of the double radio source (see also, \citealt[]{HKW2007}); 
\textbf{(d)} In a radically different scenario, superdisks are carved out as 
two galaxy cores, each containing a supermassive black hole (SMBH), merge and 
the associated pair of relativistic plasma jets undergoes a rapid precession 
during the later stage of the merger \citep[]{GB2009}. Some additional mechanisms 
proposed for the origin of superdisks are mentioned in §4.

While the different proposed scenarios seem plausible, choice of the leading contender(s) 
is contingent upon a direct 
detection of the putative superdisks, possibly in emission. With the aim to 
address this critical information deficit, we have performed a sensitive search 
for H~{\sc i} emission from four nearby FR-II \citep[Fanaroff-Riley Class II;][]{fr74} 
radio galaxies whose high-resolution maps show that the twin radio lobes have a
nearly straight edge on the side facing the host galaxy, thus
marking a strip-like radio emission gap between the lobe pair (Table \ref{tab:param}). It is worth pointing out that 
although the existence of 'superdisk' is not established through direct observations, we are not aware of any
satisfactory alternative explanation for the observational results/correlations mentioned in §1.
These four radio galaxies were selected from the first sample of ten superdisk candidates
(GKW2000), which was assembled through a visual inspection of the then available radio galaxy images.
The selected 4 galaxies are among the nearest ones in the sample.

\begin{table*}
\centering
\begin{threeparttable}
\caption{Basic parameters of the target radio galaxies (GKW2000 and references therein): (i) Name, (ii) 
Coordinates, (iii) Redshift, (iv) Distance (angular dia. distance $D_{A}$ and luminosity distance $D_{L}$), 
(v) Inferred superdisk size (minimum dia. $d$ and width $w$), and (vi) Full width at half maximum (FWHM) of the synthesized beam (see, Figure \ref{fig:noise}). See table footnote, for the references for $z$, $D_{A}$, $D_{L}$, $d$, $w$. }
\label{tab:param}
\begin{tabular}{cccccccccc}
\hline
Source & \multicolumn{2}{c}{Coordinates} & Redshift& $D_{A}$ & $D_{L}$ & \multicolumn{2}{c}{Superdisk size\tablefootnote{g}} 
& FWHM\\
 & R.A. (J2000) & Dec. (J2000) & ($z$) & (Mpc) & (Mpc) & $d$ (kpc) & $w$ (kpc) &of the synthesized beam\\
\hline
3C~98    & $03^{h}58^{m}54.0^{s}$ & $10^{h}26^{m}03^{s}$ & 0.0304\tablefootnote{a} & 126.5\tablefootnote{d} & 134.3\tablefootnote{d} & ~30 & 15 & $56.3\arcsec\times46.7\arcsec$\\
3C~192   & $08^{h}05^{m}34.9^{s}$ & $24^{h}09^{m}50^{s}$ & 0.0597\tablefootnote{b} & 239.5\tablefootnote{d} & 269.0\tablefootnote{d} & ~80 & 25& $76.6\arcsec\times46.5\arcsec$ \\
4C+32.25 & $08^{h}31^{m}27.5^{s}$ & $32^{h}19^{m}26^{s}$ & 0.0512\tablefootnote{a} & 207.5\tablefootnote{f} & 229.3\tablefootnote{f} & 150 & 52&$55.2\arcsec\times41.5\arcsec$ \\
3C~227   & $09^{h}47^{m}45.0^{s}$ & $07^{h}25^{m}20^{s}$ & 0.0858\tablefootnote{c,e} & 334.0\tablefootnote{f} & 393.8\tablefootnote{f} & ~85& 33& $59.5\arcsec\times50.7\arcsec$ \\
\hline
\end{tabular}
\begin{tablenotes}\footnotesize
\item{Refs:~[a] \citet{MASS2015}; [b] \citet{AFM2015}; [c] \citet{OKY2015}; [d] \citet{LPB1997}; [e] \citet{BBLP1992}; 
 \newline[f] \citet{WR2006}; [g] \citet{GKW2000}}
\end{tablenotes}	
 \end{threeparttable}
\end{table*}

\begin{figure*}[htp]
\centering
\begin{subfigure}{\linewidth}
\centering
\includegraphics[width=0.57\linewidth]{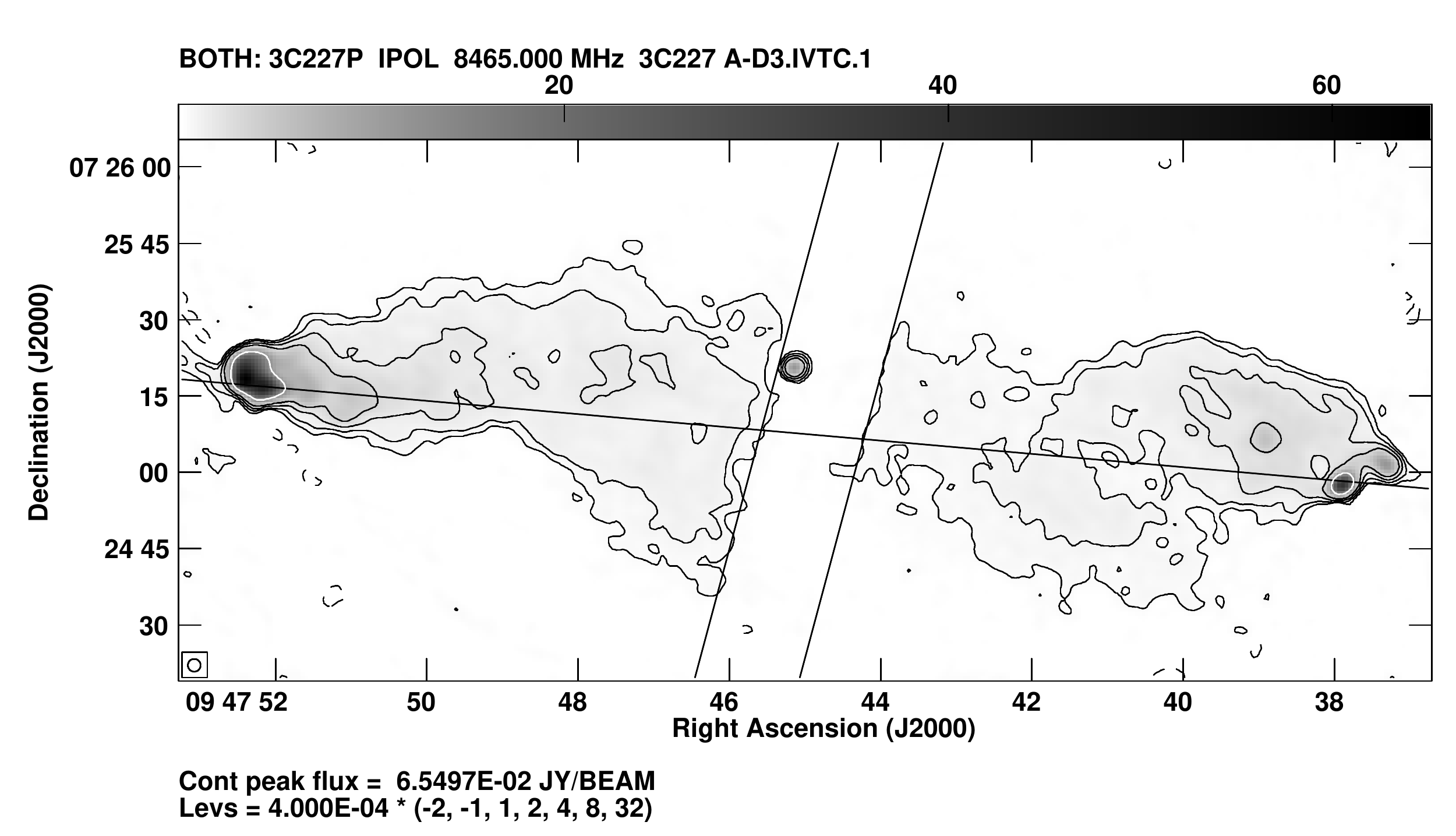}\includegraphics[width=0.47\linewidth]{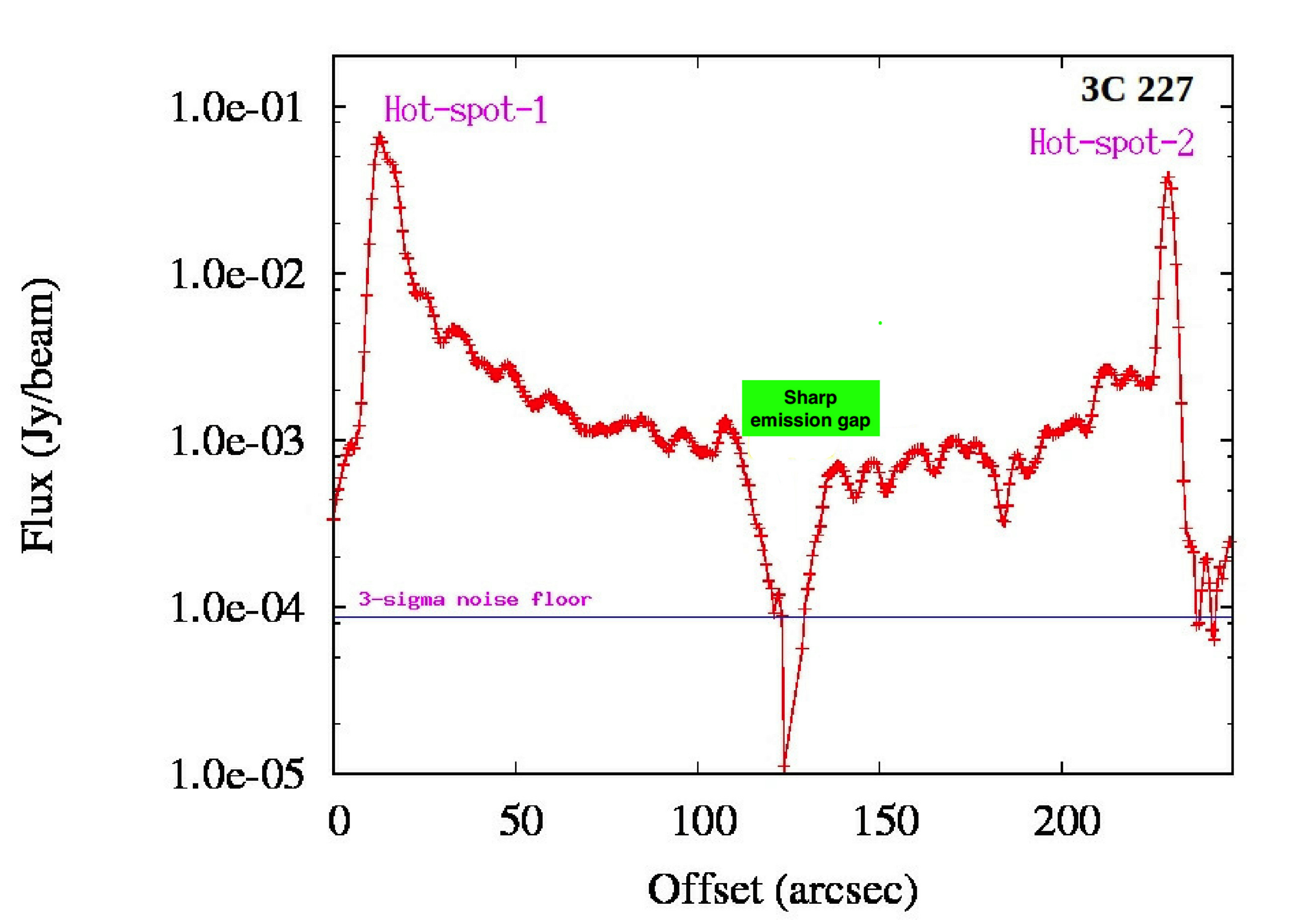}
\end{subfigure}

\caption{Left: VLA X-band (8.4 GHz) grey scale image with (-2, -1, 1, 2, 4, 8, 32) $\times$ 3$\sigma$ contours of superdisk candidate 3C 227 (with available data from \citealt[]{BBLP1992}). The two parallel lines overplotted show the region of radio emission gap. Right: 8.4 GHz radio emission profile of the same source along the axis joining the two hotspots, also overplotted on the left panel image, clearly shows the sharp emission gap.}
\label{fig:superdisk}
\end{figure*}

As an example, we have displayed in Fig \ref{fig:superdisk} for one of them (3C 227), the grey scale image and the radio brightness 
profile taken along the major radio axis. The sharply bounded central brightness depression marks the
location of the superdisk inferred to be associated with this source
(see also \citealt{BHTB1988}, \citealt{BBLP1992}, and  
``An Atlas of DRAGNs''\footnote[1]{http://www.jb.man.ac.uk/atlas/} edited by 
J. P. Leahy, A. H. Bridle \& R. G. Strom,
for high resolution radio continuum images of these sources). To the best of our knowledge, 
no targeted H I observations of superdisk candidate radio galaxies had been carried out, prior to
the ones reported here.

It may be noted that out of the ten radio galaxies listed in GKW2000, 
deemed to possess superdisks, five lie at $z < 0.1$. Here we report H~{\sc i} 
search for four of these five radio galaxies, the fifth source (3C 33) could 
not be observed due to scheduling considerations. The observational details of 
this project are provided in §2. 
In §3, we present the results. After a brief discussion, our main 
conclusions are summarized in §4.

\section{Observation and data analysis}

The four superdisk candidates in radio galaxies ($0.03 < z < 0.09$) were 
observed during January to March, 2013, with the NSF's Karl G. Jansky Very Large 
Array (VLA) in its D-configuration, yielding a baseline coverage from $0.035$ to 
$1.03$ km (VLA Project code 13A-194). The total observation time was 11 hours 
including the calibration and other overheads (total 5 hrs including 4 hrs 
on-source for 3C 227, and total 2 hrs including 1.5 hr on-source for each of 
the remaining three sources).

\begin{figure}[ht]
\includegraphics[height=\columnwidth,angle=-90]{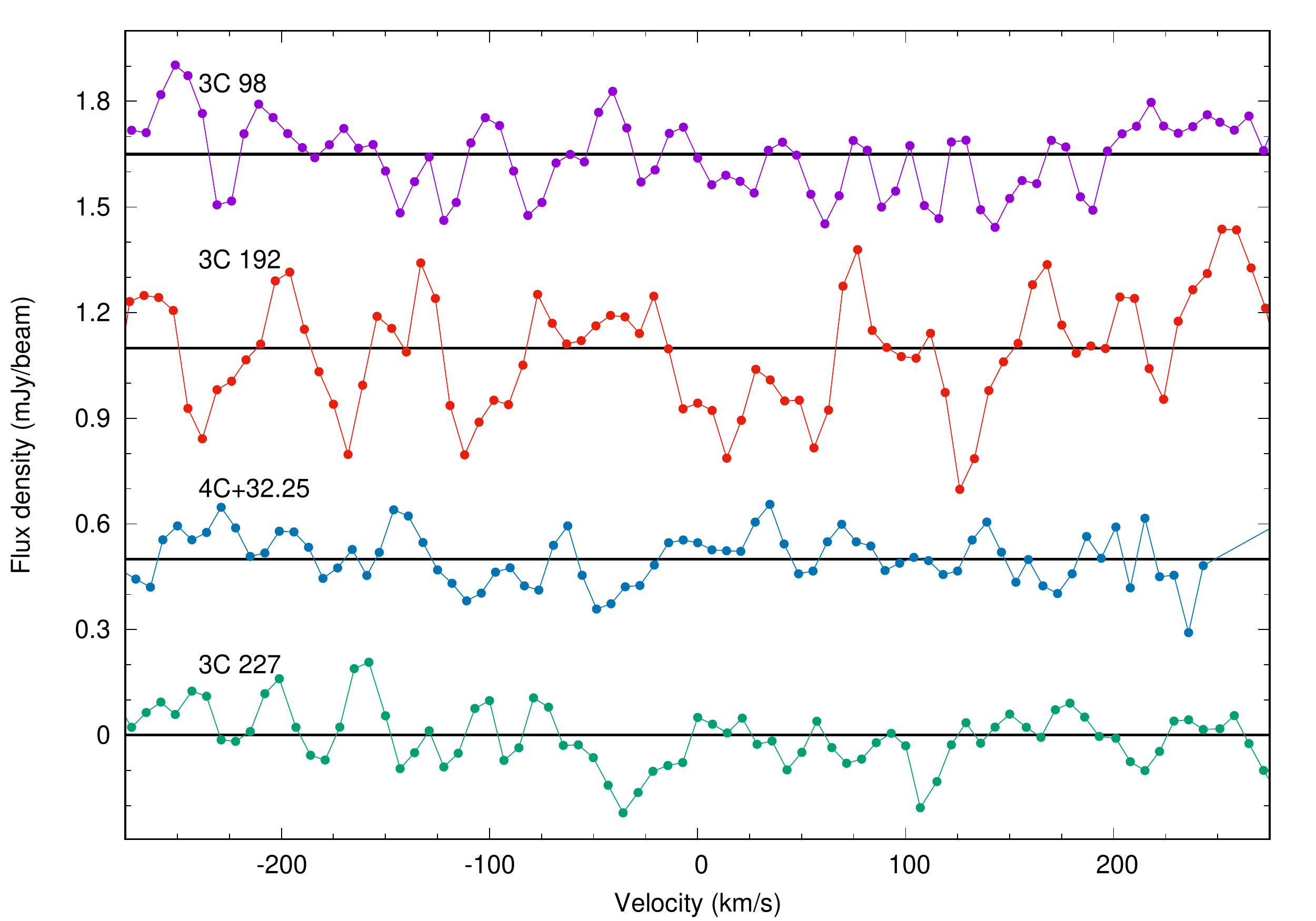}
\caption{The final continuum subtracted spectra for the four superdisk candidates, with a velocity 
resolution of $\sim 7$ \kms. The spectrum for the 3C~227 superdisk is a combination of the spectra 
derived from its observations on the two days. Note that, for clarity, the spectra have been shifted 
vertically.}
\label{fig:spectra}
\end{figure}

To achieve a reasonable sensitivity and uv-coverage, the 
observations were carried out with a bandwidth of 16 MHz centered at the 
redshifted H~{\sc i} frequency and using 512 spectral channels. This yielded an 
overall velocity width of $\sim 3500$ \kms at a spectral resolution of $\sim 7$ \kms.
The relatively high spectral resolution enabled a more effective RFI flagging. 
We opted for a large velocity coverage considering the total lack of {\it a priori} 
information about the width of the putative H~{\sc i} emission line. The 512 
channels provided enough ``line-free'' (and RFI-free) channels to be able to 
properly subtract  the continuum and thus establish a flat spectral baseline. 
The data reduction  was done in the Common Astronomy Software Applications (CASA) 
package version - 4.6.0 \citep[]{CASA2007} and the Astronomical Image Processing 
System (AIPS)  version - 31DEC17 \citep[]{gustav96} following standard procedures. 
As a consistency check, the initial flagging and calibration for one of the targets 
(namely, 3C 98) were carried out both manually and using the VLA Scripted 
Calibration Pipeline version - 1.3.8 \citep[]{CASA2007}. Subsequently, for 
flagging and calibration of the rest of the data, we only used the pipeline. 
Additional data flagging, initial imaging and self-calibration of the continuum 
data, and the final continuum imaging were done using AIPS. The  AIPS task UVSUB 
was then used to subtract the continuum, CVEL to accurately convert the 
observed frequency to Heliocentric velocity, and thereafter, IMAGR was run on the 
residual data to make image cubes at various spectral resolutions. Residual 
spectral baseline was removed by using the task IMLIN on the spectral data 
cube. Finally, we visually inspected the cube for any emission signatures and to extract the
spectrum of the central radio emission gap. For 3C 227, the data 
from the two observing sessions were analyzed separately and then combined with inverse 
variance weighting at the final stage subsequent to the continuum subtraction.

\begin{figure*}[ht]
\includegraphics[width=0.5\columnwidth]{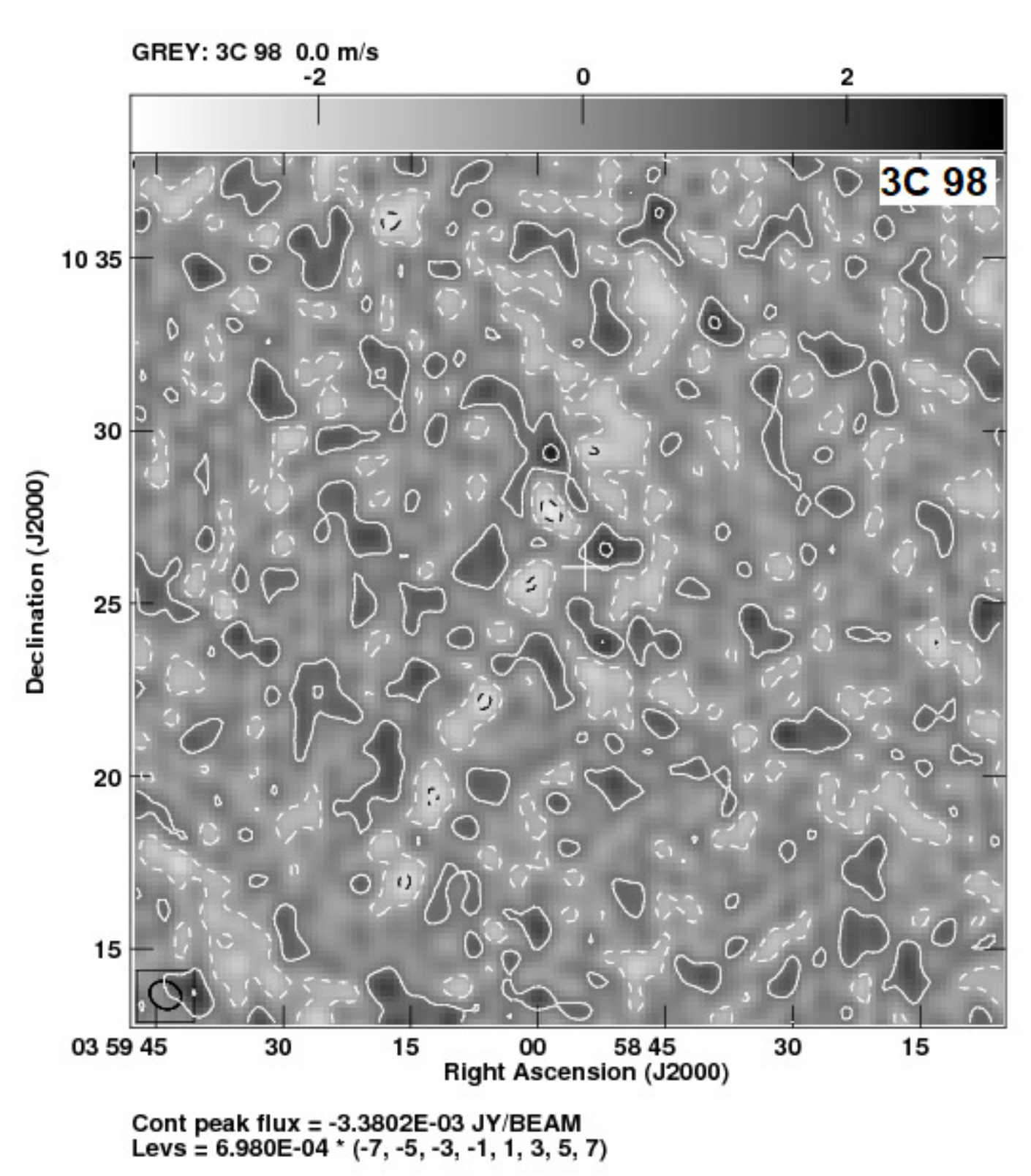}\includegraphics[width=0.5\columnwidth]{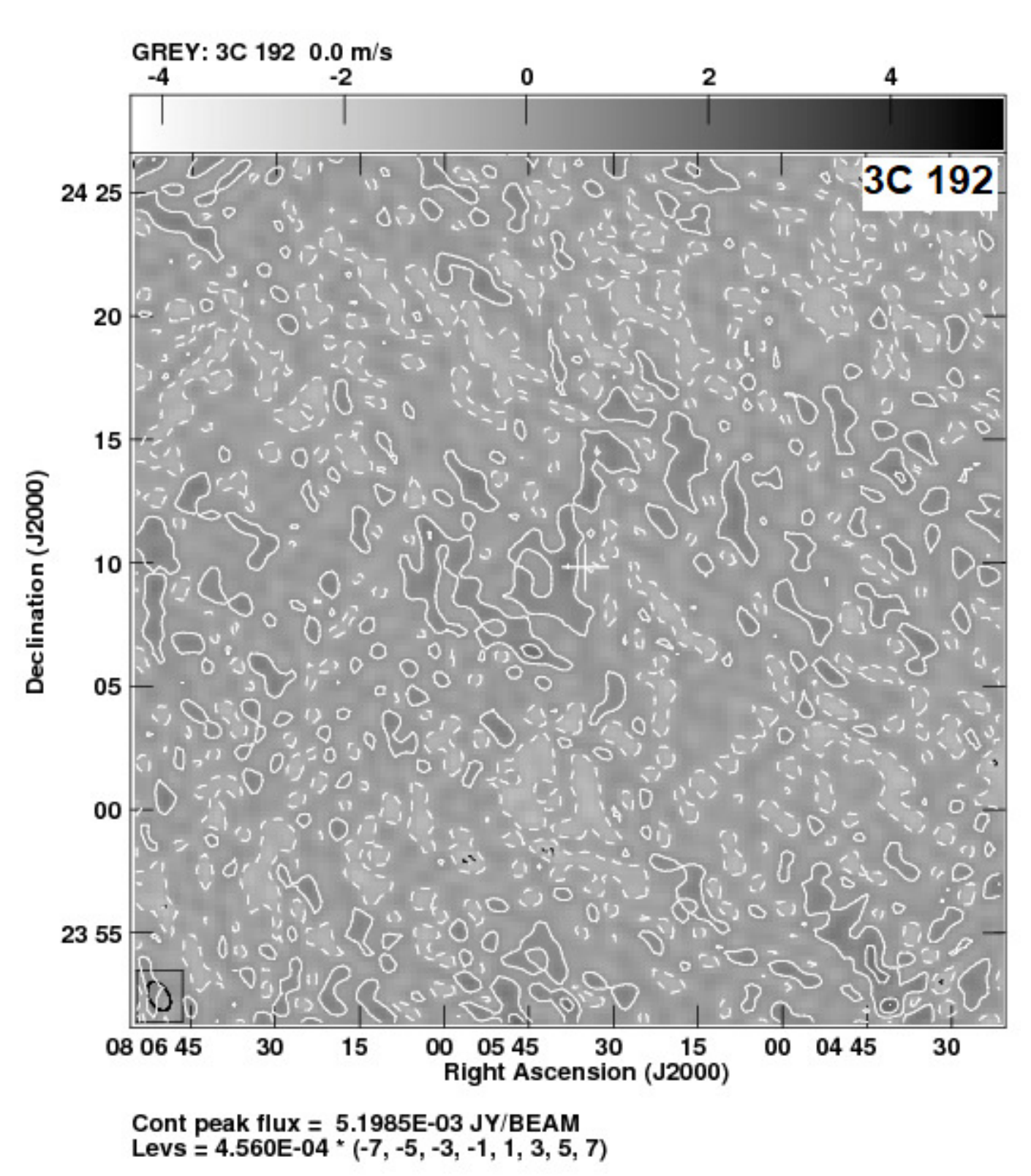}\\
\includegraphics[width=0.5\columnwidth]{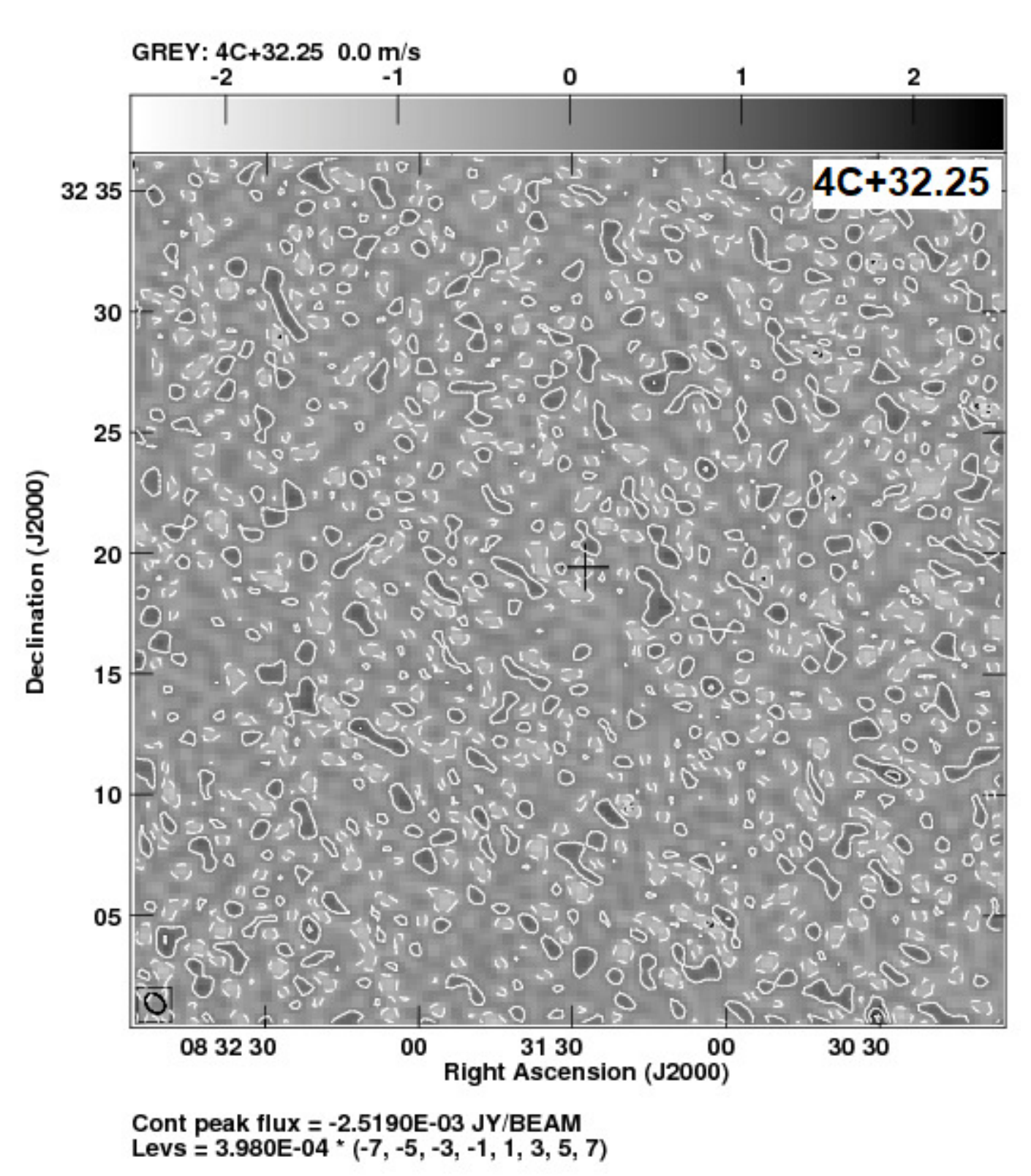}\includegraphics[width=0.5\columnwidth]{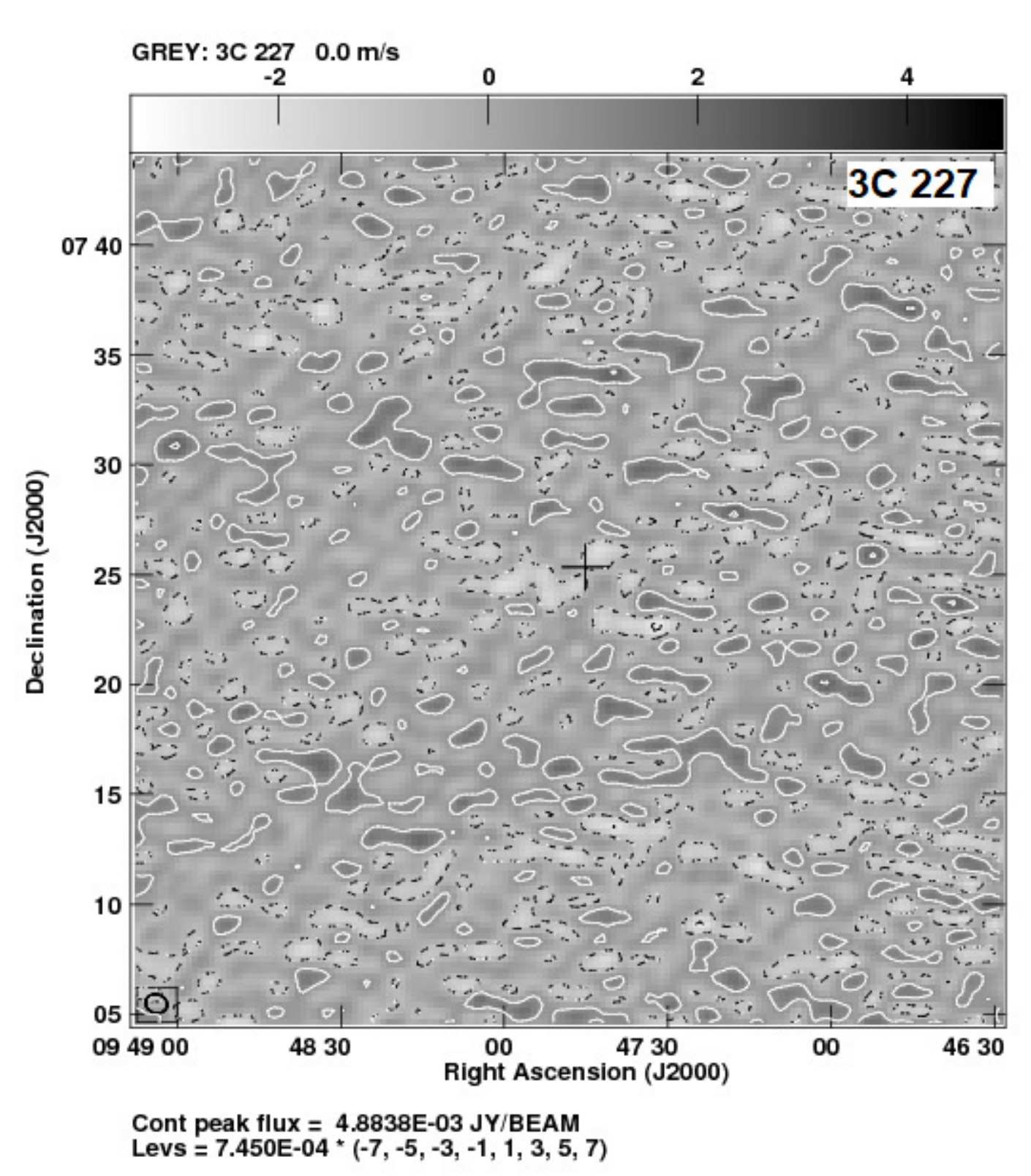}
\caption{The final continuum subtracted images with $\pm$(1,3,5,7)$\sigma$ contours overplotted, 
at the redshifted H~{\sc i} 21 cm frequency. The signal has been averaged over $100$ \kms velocity bin. 
The brightness scale is indicated at the top of each plot. The central cross indicates the position of 
the nucleus of the radio source. The full width at half maximum (FWHM) of the synthesized beam, shown 
within the inset in each image, are $56.3\arcsec\times46.7\arcsec$ (3C 98), $76.6\arcsec\times46.5\arcsec$ 
(3C 192), $55.2\arcsec\times41.5\arcsec$ (4C+32.25), and $59.5\arcsec\times50.7\arcsec$ (3C 227).}
\label{fig:noise}
\end{figure*}

\section{Results}

\begin{figure}[ht]
\includegraphics[width=\columnwidth]{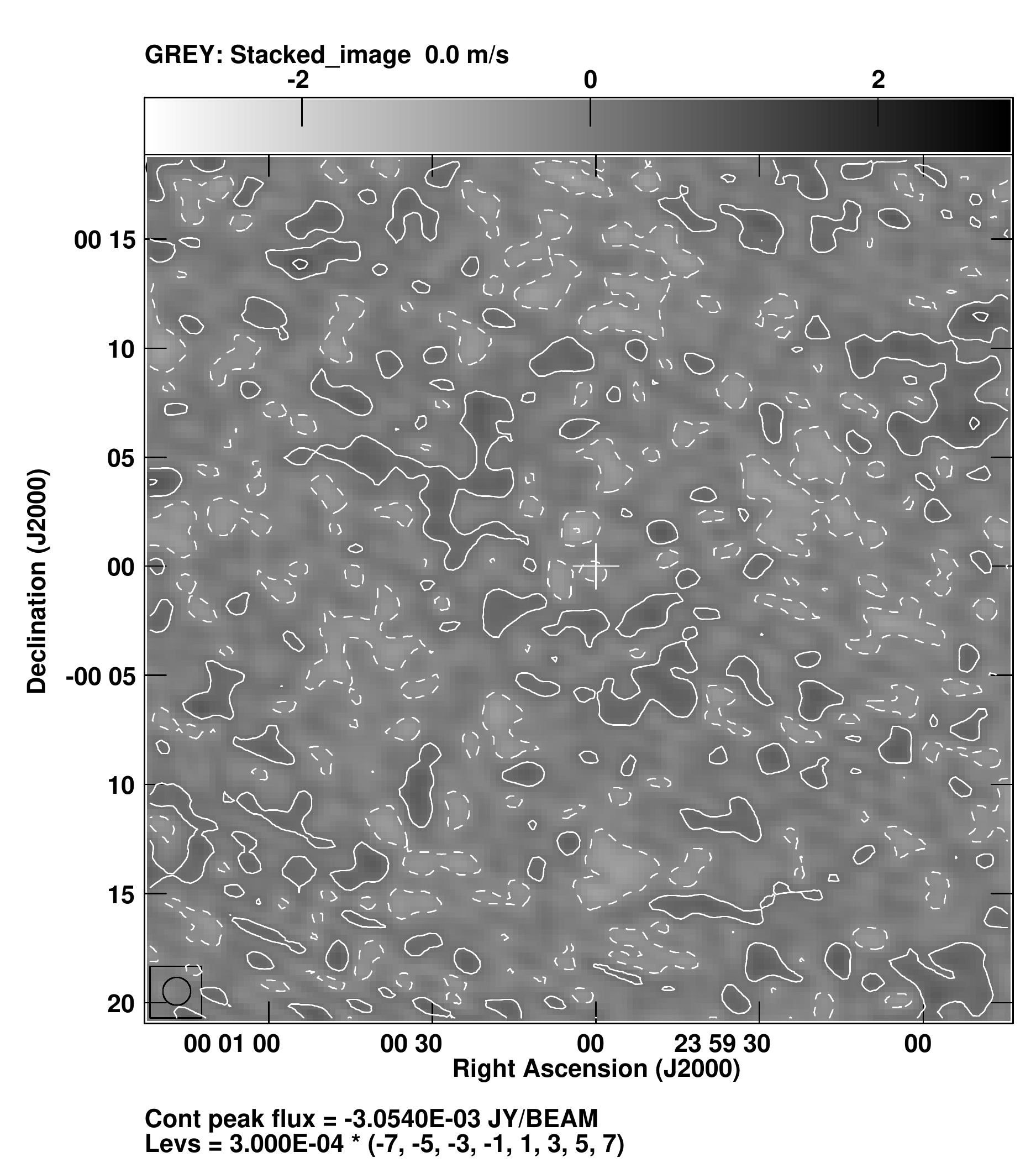}
\caption{Stacked H~{\sc i} image of the four superdisk candidates, each smoothed to $76.6\arcsec\times76.6\arcsec$ beam 
and then combined with inverse variance weighting, after aligning at the redshifted H~{\sc i} 21 cm frequency and averaging 
over $100$ \kms velocity bin. The grey scale image is overplotted with $\pm$(1,3,5,7)$\sigma$ contours of H~{\sc i} signal.  
The scale is indicated at the top. The rms noise in the combined map is 0.3 mJy/beam and the inset shows the FWHM of the 
restoring synthesized beam.} 
\label{fig:comb}
\end{figure}

Despite a fairly high sensitivity, the present VLA observations have not 
yielded a positive 
signal of H~{\sc i} 21 cm emission from any of the four superdisk candidates. 
Figure \ref{fig:spectra} displays the spectra extracted at their locations, for 
a $\pm 300$ \kms velocity range, at the original spectral resolution. Even when 
the spectral cubes were smoothed to a resolution of $\sim 100$ \kms, no 
significant signal of H~{\sc i} emission was detected for any of the four 
targets (Figure \ref{fig:noise}). The cross in each panel shows the central
position of the emission gap at which the spectrum was extracted. We note that 
even after a careful flagging, calibration and imaging there are some residual 
low-level artifacts, probably due to uv-coverage limitations or residual 
RFI in some of the images. These, however, do not significantly affect the central 
region from where the spectra have been extracted. We set the 3-sigma noise as the 
upper limit to H~{\sc i} mass under the assumptions of optically thin line emission 
\citep[]{DRAINE2011} and a line width of 100 \kms. We then used the estimated size 
of the putative superdisk (Table \ref{tab:param}) to compute the corresponding 
upper limit to H~{\sc i} number density and (averaged) H~{\sc i} column
density (by a spatial averaging performed over the inferred superdisk region). 
These results are summarized in Table \ref{tab:results}. It should be remembered 
that the size estimate for each superdisk is quite approximate, as the superdisk may extend even beyond
the lateral extent of the radio emission gap. In each case, the spatial resolutions of the present 
observations is such that the synthesized beam size is very similar to the size of 
the emission gap and hence the beam should encompass much of the putative superdisk.
Lastly, in a bid to further enhance the sensitivity, we have combined the final four H~{\sc i} 
images, all smoothed to a common resolution of $76.6\arcsec\times76.6\arcsec$ (the largest 
among the synthesized beams for the four sources), with inverse variance weighting. 
However, as seen from Figure \ref{fig:comb}, even this stacking of the H~{\sc i} spectra of 
the four superdisk candidates has not yielded a significant H~{\sc i} signal.

\section{Summary and discussion}

The present VLA H~{\sc i} imaging observations have provided the first 
information on the H~{\sc i} content of the putative superdisks, For this, we 
have carried out H~{\sc i} imaging of a sample of four nearby radio galaxies in 
which a superdisk has been inferred on the basis of high-resolution radio 
continuum maps (Table \ref{tab:param};  §1). Although the derived upper limits to 
H~{\sc i} mass ($10^{8}-10^{9}$ $M_{\sun}$) for these superdisks are similar 
to the H~{\sc i} content of the Milky Way, the corresponding average H~{\sc i} number 
densities have upper limits of only $10^{-4} - 10^{-3}$ atoms per cm$^3$, 
thanks to the immense volumes of the inferred superdisks. 
We reiterate that the superdisk dimensions adopted here are approximate as they refer only to 
the region of the strip-like radio emission gap (or brightness depression) seen between 
the radio lobe pair; in reality the superdisk may extend well beyond the lobe 
width, resulting in even tighter upper limits on H~{\sc i} number 
densities. Considering this, upper limits on the average H~{\sc i} \textit{column 
density} might be more 
meaningful (Table \ref{tab:results}). Thus, while extension of H~{\sc i} 
searches to more superdisk candidates and to deeper levels would be very 
desirable, the present study does point to a scarcity of H~{\sc i} in 
superdisks, the highly suggestive evidence for dust in superdisks 
notwithstanding (§1). Here it may be mentioned that the presence of dust
or dusty globules/filaments, 
per se, does not rule out a superdisk medium dominated by a warm/hot gas. For 
instance, dust emission has been convincingly detected in the central parts of the 
X-ray emitting clusters of galaxies \citep[e.g.,][]{YK2005,MG2005,TYYH2009}.
It is also important to clarify that our observations do not exclude the existence of 
a small, kpc-scale H~{\sc i } disk in these galaxies. For comparison, the typical 
H~{\sc i} column density of such disks can be of the order of $0.6-3.1\times10^{19}$ 
atoms per cm$^2$ \citep[]{SERRA2012}, which would remain undetected with the surface-brightness
sensitivity of the present wide-beam images targeted at much larger H~{\sc i} structures (i.e., superdisks)
that are postulated to explain the observed huge radio emission gaps. The present observations 
only imply that the volume density of H~{\sc i} in the putative superdisks must be {\it at least} 
an order of magnitude lower than that estimated for the kpc-scale H~{\sc i} disks 
seen in some early-type galaxies \citep[e.g.,][]{SERRA2012}.


\begin{table}
\centering
\caption{The 3$\sigma$ {\it upper limits} to the H~{\sc i} number density, column density and mass of the putative superdisks 
in the four observed radio galaxies, assuming a velocity width of $\sim100$ \kms. Note that the listed RMS noise values refer 
to the velocity width resulting from spectral smoothing over 15 channels, in order to obtain a velocity resolution of 
100 \kms. These are $\sim 5 - 10$ times above the confusion level in the D configuration for VLA L-band observations 
(74 $\mu$Jy/beam).}
\label{tab:results}
\begin{tabular}{lcccc}
\hline
Source & RMS noise  &  $<$n(H~{\sc i})$>$  & $<$N(H~{\sc i})$>$ & M(H~{\sc i})\\
   & ($\mu$Jy/beam) & (cm$^{-3}$) & (cm$^{-2}$) & (M$_{\sun}$)\\
\hline
3C~98    & $698$ & $1.6\times10^{-3}$ & $1.5\times10^{20}$ & $4.2\times10^{8}$\\
3C~192   & $456$ & $3.5\times10^{-4}$ & $8.7\times10^{19}$ & $1.1\times10^{9}$\\
4C+32.25 & $398$ & $2.4\times10^{-4}$ & $1.1\times10^{20}$ & $5.4\times10^{9}$\\
3C~227   & $745$ & $7.1\times10^{-4}$ & $1.9\times10^{20}$ & $3.3\times10^{9}$\\
\hline
\end{tabular}	 
\end{table}

One class of gaseous structures that might have some bearing on the superdisk 
scenario is the so called ``gas belts'' \citep[][and references 
therein]{MWB2013}. These X-ray detected structures have been identified in a 
few radio galaxies located in group/cluster-like environments, namely  3C 35 
\citep[]{MWB2013}, 3C 285 \citep[]{HKW2007}, 3C 386 \citep[]{DW2016}, and 3C 
442A \citep[]{HKW2007, WBKH2007}. Broadly, the belt-like structures of hot 
gas associated with these radio galaxies are seen to overlap with the radio  
gap/depression between their 
twin radio lobes and are oriented roughly orthogonally to the axis defined 
by the radio lobe pair. No consensus has yet emerged about their formation mechanism. 
\citet[]{HKW2007} have argued that the central X-ray ridge, i.e., the gas belt, 
in 3C 285 is a ridge of thermal plasma that pre-existed the commencement of the 
radio activity, an inference also reached in \citet[]{GKW2009}, 
from the observed greater lobe-length symmetry when measured relative to the mid-plane 
of the superdisk, rather than when referenced to the host galaxy. On the other 
hand, for the case of 3C 442A, \citet[]{WBKH2007} have argued that the gas belt 
is the hot gas being stripped from the merging galaxies and is getting aligned 
into a belt shape as it pushes the twin radio lobes straddling it \citep[see also,]
[for a scenario involving a dynamical interaction of the hot thermal gaseous halo with the radio lobe pair]{GKWJ2007}. 
A yet another explanation, advanced by \citet[]{MWB2013} 
for the case of 3C 35, posits that its gas belt has formed out of 
the gas belonging to the fossil galaxy group but driven outward due to the 
expansion of the twin radio lobes. More recently, a temperature gradient has 
been detected along the projected 53 kpc length of the gas belt observed in 3C 
386, with a cooler gas core ($T \sim 0.73 \pm 0.09$ keV) found closer to the 
host elliptical and the hotter ($T \sim 1.72 \pm 0.46$ keV) outer portion of 
the belt interpreted as the group's atmosphere which is undergoing an inward 
flow triggered by the buoyant rise of the twin radio lobes through the 
circumgalactic medium \citep[]{DW2016}. While all these physical processes may be 
operating to varying degrees, the striking straightness of the lobes' brightness 
contours marking the edges of the central radio emission gap/depression which we 
identify as superdisk, continues to pose a challenge and its origin remains to be properly understood. 
Perhaps even more enigmatic are the several radio galaxies, highlighted in 
\citet{GKW2009}, where the host elliptical galaxy is located nearly at one 
edge of the strip-like central radio depression (i.e., superdisk), or even 
{\it outside} it, i.e., within a radio lobe  (this puzzling extreme situation is witnessed in 
the radio galaxy 3C 16, e.g., \citealt{HH1998}). Since superdisks are likely to 
contain both ionized and dusty cool gas, a multi-pronged observational follow-up 
will be important, including the
possibility of H~{\sc i} search through sensitive absorption measurements against 
any available background radio source.

\begin{acknowledgements}
We are grateful to Eric Greisen, Sanjay Bhatnagar, Arnab Rai Choudhuri and Prateek Sharma for useful 
discussions. This research has made use of NASA's Astrophysics Data System. The results reported here 
are based on observations made with the Karl G. Jansky Very Large Array of the National Radio Astronomy 
Observatory. The National Radio Astronomy Observatory is a facility of the National Science Foundation 
operated under cooperative agreement by Associated Universities, Inc. Most of the work was done when AA was at Indian Institute of Science as a undergraduate student. NR acknowledges support from the Infosys Foundation through the Young Investigator grant.
G-K thanks  the Alexander von-Humboldt Foundation for financial support and Prof. A. Zensus, Director, Max-Planck-Institut 
f{\"u}r Radioastronomie, Bonn for the local hospitality.
\end{acknowledgements}

\bibliographystyle{raa}
\bibliography{ms_raa2018_0287_r1} 
\end{document}